\documentclass[aps,prl,reprint,groupedaddress,twocolumn]{revtex4-1}
\usepackage{graphicx}% Include figure files
\usepackage{dcolumn}% Align table columns on decimal point
\usepackage{color}
\usepackage{tikz}
\usetikzlibrary{shapes}
\usepackage{gensymb}
\usepackage{epstopdf}
\begin{document}
\preprint{AIP/123-QED}
\title[MAGNETIC STRUCTURES]{Sudden Magnetization Drops in S/F bilayered Structures}
\vspace{1cm}
\author{Edgar J. Pati\~no}
\affiliation{School of Physical Sciences and Nanotechnology, Yachay Tech University, 100119 Urcuqu\'i, Ecuador}
\affiliation{Departamento de F{\'i}sica, Grupo de F\'isica de la Materia Condensada, Universidad de los Andes, Carrera 1 No. 18A-12, A.A. 4976-12340, Bogot{\'a}, Colombia.}
\author{M. G. Blamire}
\affiliation{Department of Material Science and Metallurgy, University of Cambridge, Pembroke Street, Cambridge CB2 3QZ, United Kingdom.}
\date{\today}
\begin{abstract}

We report in plane magnetization measurements on Nb/Co bilayers using a thin superconductor Nb$\sim25$ nm and a thick ferromagnet Co$\sim40-55$ nm where Bloch Domain Walls (BDW) are energetically favorable. Here sharp drops in the magnetization are explained as direct consequence of BDW stray fields that induce vortices in the out of plane direction. These are generated at small fields around the coercive field of the ferromagnet and removed at large fields of the order of $H_{c2}$. These two well distinctive magnetic states of the superconductor, with or without out of plane vortices, is the same mechanism used to explain the drops in critical currents reported in [E. J. Pati\~{n}o et. al., The European Physical Journal B 68, 73 (2009)].
\end{abstract}
\pacs{74.25.Ha, 74.45.+c, 75.30.Et}
%
% Classification Scheme.
% %%\keywords{Superconductor, Ferromagnet, domain walls, BDW stray fields, vortices }%Use showkeys class option if keyword
%display desired
%
\maketitle

\textbf{I. INTRODUCTION}\\\\

During the past 15 years Superconductor (S)/Ferromagnet (F) multi-layered structures have been a subject of intensive research. The interests in S/F heterostructures originated from the new physical properties that arise due to stray field  \cite{1Patino,16Kobayashi,17Kobayashi,27Lange,Silhanek,Vlasko,Carapella,Perez,Visani,Gomez} and proximity effects \cite{2Xia,3Salikhov,4Salikhov,5Jiang,6Lazar,7Garifullin,8Zdravkov,9Zhu, 10Ryazanov,11Kontos,9Zhu,12Robinson,13Khaire,14Leksin,15Patino}
Furthermore more recently induced triplet pairs (with integer spin) in S/F structures have
%has recently
opened the window for applications in superconducting spintronic devices \cite{Banerjee}.

Most of the  previous works dealt with transport measurements of S/F structures and only a few \cite{15Patino,16Kobayashi,Chacon,Chacon2} investigated  magnetic properties of the superconductor under the influence of the ferromagnet.
Indeed, as recent experiments suggest the magnetic properties of F/S/F spin valve structures may strongly be affected by proximity effect leading to strong variations of the superconductor's magnetization \cite{15Patino}.  Here using a thin superconductor it was possible to squeeze
vortices between two ferromagnets allowing a direct control the vortex orientation by ferromagnets' relative magnetic orientation leading to
vortex flipping.
%or coercive field reduction
%driven .
This effect replaces vortex annihilation processes, in conventional
superconductors, thus it may find applications in
reducing AC losses and preventing flux jumps
\cite{15clem, 25Evetts}.

All of the effects that come as consequence of the proximity effect, have been studied for small thicknesses (typically a few nanometres) of the ferromagnet.  When strong or thick ferromagnets are used the effects of stray fields may be observable in the experiments.  There are two possible stray field contributions, arising either from magnetic poles or domain walls. These generate flux which pierces the superconductor. \\
When a S/F structure is exposed to externally applied field the ferromagnet goes from multidomain to single domain state
leading to significant dipole stray fields. Evidence of these has been reported for Co/Nb/Co trilayer structures with thick Co and Nb layers measured by magnetization and critical current measurements \cite{15Patino,16Kobayashi,17Kobayashi}.\\

%%%%%%%%%%%%%%%%%%%%%%%%%%%%%%%%%%%%%%%%%%%%%%%
On the other hand when the ferromagnet is in multi-domain state stray fields from domain walls appear.
Depending on the thickness of the ferromagnet domain walls can be classified into two types, Neel and Bloch.
Neel domain walls are understood as the separation between domains of opposite polarity,  formed by magnetic dipoles that rotate
within the plane of the structure. On the other hand in Bloch domain walls the magnetic dipoles rotate perpendicular to the plane, leading to
magnetization and strong stray fields in the out of plane direction  (bottom right inset Fig.\ref {fig3}a).
%
%%%%%%%%%%%%%%%%%%%%%%%%%%%%%%%%%%%%%%%%%
%
The thickness transition between these two types is determined by demagnetizing fields which means that it becomes increasingly energetically favourable for domain walls to mutate from out of plane to in-plane direction as films become thinner. This has been shown in experiments in Co, Permalloy (Py) and Ni samples \cite{Carapella,1Patino,18Hsieh,19Lohndorf,20Methfessel} where the transition between these two type of domain walls occurs at around 30 nm for Co and 20 nm for Ni.\\

In experiments where the usual proximity effects (due to singlet pairs) were studied \cite{5Jiang,7Garifullin,8Zdravkov,9Zhu,10Ryazanov,11Kontos,21Larkin}
the chosen ferromagnet thickness was of the order of the ferromagnetic coherence length $\xi_F$; in this range one usually expects domain walls to lie in the plane. For thick ferromagnets Bloch domain walls (BDW) dominate and may generate vortices in the superconductor in the out of plane direction \cite{22Ryazanov} affecting its critical current and magnetoresistance  \cite{Carapella,1Patino,17Kobayashi}.
Here, given that the thickness of the ferromagnet was much greater than $\xi_F$, effects due to Bloch domain walls (BDW) prevail over
the  so-called  Fulde-Ferrel-Larkin-Ovchinnilov (FFLO) state, effects which have been damped at the ferromagnet coherence length scale.
\\

The present work can be considered as an extension of our previous paper on critical current measurements \cite{1Patino} of Nb/Co bilayers.  Here dips of the critical current and peaks in the magnetoresistance were observed between coercive and saturation fields. These were accompanied by vortex flow onset below the normal critical current value indicating that vortex motion was involved. Because the measurements were made in a Lorentz force-free configuration (i.e. current parallel to the field) this vortex motion indicated the presence of vortices perpendicular to the plane which caused this voltage onset. These dips were explained as result of a sudden vortex-antivortex 2D lattice formation (perpendicular to plane) leading to a reduction of the critical current. These are the result of Bloch lines formed with strong antiparallel stray fields at the domain walls. Increasing the field further and fully magnetizing the sample (where BDW are absent) resulted in an increase in the critical current back to its original value. Similar results were independently found in Co/Nb/Co trilayers \cite{Carapella}.\\

In contrast, in this paper we performed in-plane magnetization measurements on Nb/Co continuous bilayer films. This allowed us to investigate the effect of the Co magnetization on the superconductor's magnetization. The ferromagnet thickness was chosen to be larger than $\xi_F$, to avoid FFLO effects, and thick enough to allow Bloch domain wall effects. Contrary to \cite{1Patino} in this investigation there are no external currents and therefore Lorentz forces are absent.\\

As in our previous investigations, the superconductor thickness was kept small $d_s$$\sim$$2$$\xi_{GL}$ (the Ginzburg Landau coherence length), as it is well known that as the superconductor gets thinner, $T_{c}$ and  $H_{c}$ falls below their bulk values \cite{23Bell}.  This makes these structures more susceptible to vortex penetration. As we shall see, the effect of vortex injection from the Co layer leads to a sudden drop in magnetization of the superconducting layer.  A reference Nb/Co bilayer with a thin Co layer of ~ 3 nm was also studied \cite{15Patino} to compare our results; this eliminated effects arising from Bloch domain walls while keeping the usual proximity effect. Furthermore this Nb/Co bilayer is a good reference sample for obtaining characteristic parameters of the ultra-thin Nb film.\\

\textbf{II. EXPERIMENT}\\

The Nb/Co bilayers were grown using a UHV DC-magnetron sputtering system in a chamber cooled to $-100 \degree C$ using liquid nitrogen as reported in \cite{1Patino}. The base pressure prior to Ar insertion was less than $2\times 10^{-9}$ Torr. The partial oxygen base pressure in the chamber
was less than $0.7 \times 10^{-11}$ Torr as measured using a mass spectrometer . Films were deposited from $99.95 \%$ pure sputtering targets using an Ar pressure of $3.7\times 10^{-3}$ Torr, in an in-plane magnetic field of approximately $400$ Oe. The bilayer sample was grown on Si ($100$) substrates, with a Nb thickness $d_{Nb} = 25$ nm. The list of studied samples in the present work is summarized in Table I. Here the numbers in parenthesis indicates the thickness of each layer in nm.\\

\begin{table}[h]
\centering
\begin{tabular}{| c | c | c |}
\hline
\hline
Sample & d$_{Co}$ & Magnetization Drops \\ \hline
		       &    &     \\
\ NbCo(3)    \ & 3  & No  \\
\ NbCo(40)   \ & 40 & Yes \\
\ NbCo(42)-1 \ & 42 & Yes \\
\ NbCo(42)-2 \ & 42 & Yes \\
\ NbCo(42)-3 \ & 42 & Yes \\
\ NbCo(42)-4 \ & 42 & Yes \\
\ NbCo(45)   \ & 45 & Yes \\
\ NbCo(55)   \ & 55 & Yes \\
		       &    &     \\
\hline
\hline
\end{tabular}
\caption{List of studied samples in the present work. First column gives the sample label, second column the thickness of the Co layer in nm and third column indicates whether or not magnetization drops where observed in such sample.}
\label{table:1}
\end{table}

Magnetization measurements were performed with the field parallel to the film surface using a Quantum Design (MPMS) superconducting quantum interference device (SQUID) magnetometer, and an Oxford Instruments vibrating sample magnetometer. The SQUID measurements were carried out in “DC and no overshoot” mode on samples with dimensions of $5 \times 5$ mm$^2$. Before starting the measurements the magnet was cycled down from $6000$ Oe to zero using an oscillating field sequence in order to remove any trapped flux in the superconducting magnet. All SQUID and VSM measurements were performed after zero field cooling (ZFC) the samples.\\

\textbf{I) REFERENCE SAMPLE}\\

To obtain the magnetic response and basic transport properties solely due to proximity effect we studied, in Ref \cite{15Patino}, a Nb($25$)/Co($3$) reference sample. Here a superconducting transition temperature  $T_{c}$ $\sim$ 6.4 K and mean free path $\sim$ 3 nm,  were obtained from resistivity measurements.
% $T_{c}$ was defined as the value where the resistance drops to half of the resistance above transition giving a  transition width of $0.1$ K.
%%%%%%%%%%
Using the expressions; $\xi_S$$=$$(\hbar D_S/2\pi k_b T_c)^{1/2}$ and $\xi_{GL}(0)$$=$$\pi$$\xi_S$$/$$2$;
The superconducting coherence length $\xi_S \sim $7 nm and $\xi_{GL} \sim$ 11 nm were extracted from these measurements.
Here the diffusion coefficient  $D_S$ was calculated using a Fermi velocity $v_F = 2.77 \times 10^5$ m/s $ $    \cite{24Weber}.
This corresponds to the vortex core radius
giving an estimate of the vortex core diameter $2\xi_{GL}$~$\sim 22$ nm slightly smaller than the Nb layer thickness.\\
The ferromagnetic coherence length, that corresponds to the traveled distance of Cooper pairs into the ferromagnet,  $\xi_{Co}$ $\sim$ 0.3 nm was also found. \\

\begin{figure}[h!]
\includegraphics[scale=0.38]{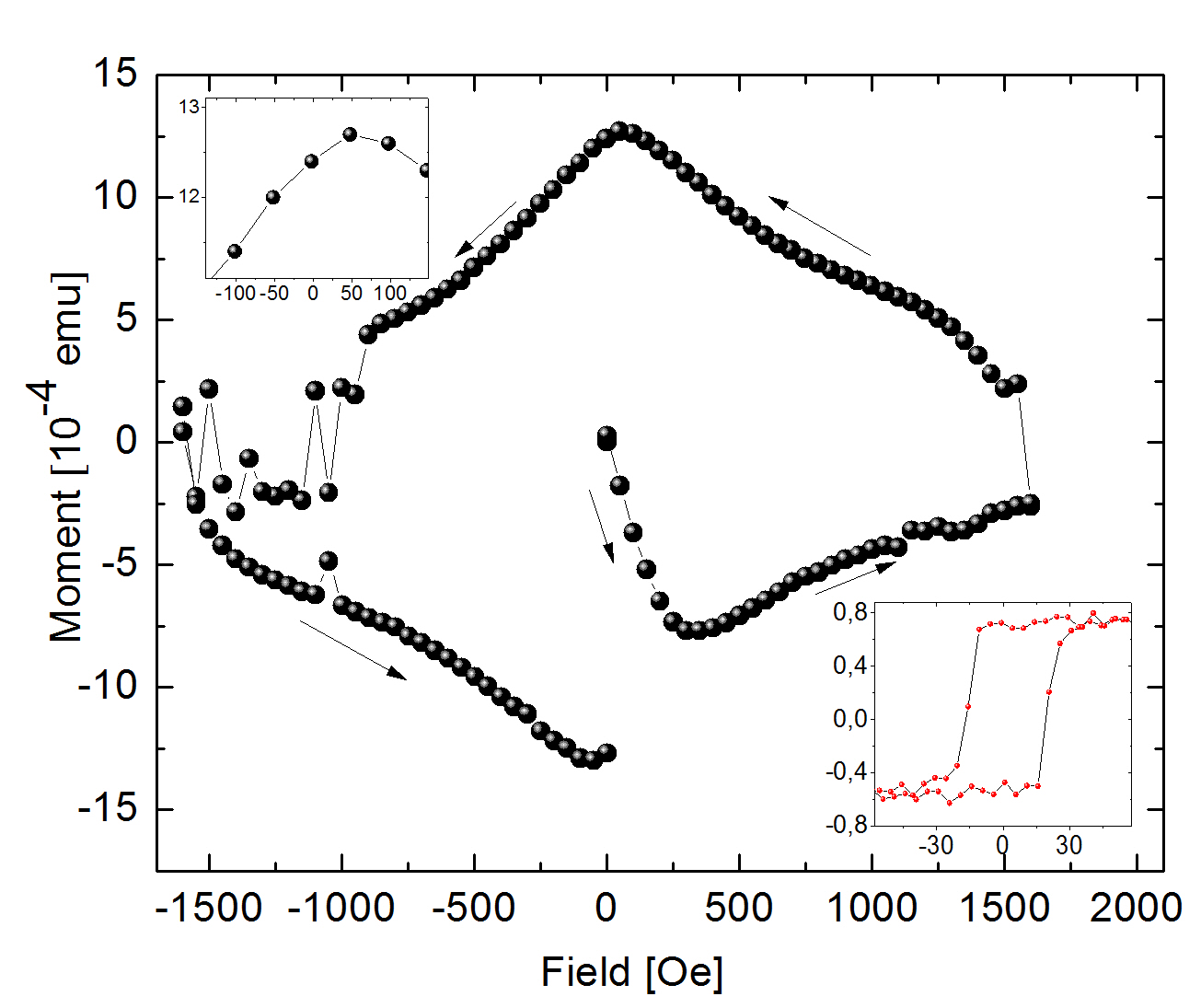}
\caption{\label{fig1} Hysteresis loop of a Nb($25$ nm)/Co($3$ nm) taken at $4.2$ K \cite{15Patino}. Upper inset shows peak position shift of $50$ Oe relative to the zero position. Lower inset shows hysteresis Loop at room temperature.}
\end{figure}

The magnetization measurements of  Nb($25$)/Co($3$) bilayers taken at room temperature showed
coercive fields of about $20$ Oe (lower inset to Fig. \ref{fig1}). When cooled down, and the characteristic magnetization loop of a type II superconductor was obtained at $4.2$ K (Fig. \ref{fig1}). A value of $H_{c1||} = 100$ Oe at $4.2$ K is found from the hysteresis loop where the magnetization slope loses its Meissner linear behaviour.\\

In this curve the three characteristic vortex processes of a type II superconductor were observed:  vortex penetration, vortex pinning and vortex creation-annihilation.
Due to spontaneous vortex creation annihilation and heat release processes \cite{25Evetts}  flux jumps were observed as usual for the in-plane measurement configuration. A shift of the central peak position near zero field towards a positive field of  $\sim 50$ Oe (upper inset Fig. \ref{fig1}). This is an indication of the presence of dipole stray fields from the magnetic layers but still smaller than $H_{c1}$.\cite{16Kobayashi}\\

\textbf{II. Nb($25$ nm)/Co(X nm) BILAYERS}\\

\textbf{MAGNETIZATION MEASUREMENTS ON Nb($25$)/Co(X) BILAYERS}\\

For this part of the investigation we proceeded with a thick ferromagnet with a Co thickness $d_{Co}$ $\geq$ $40$ nm. This is more than one order of magnitude larger than the ferromagnetic coherence length $\xi_{Co}$ $\sim$ $0.3 $nm in the dirty limit, so we don’t expect any of the phenomena related to proximity effect to be observable. On the other hand a large thickness of the ferromagnet results in domain walls being of the Bloch type which lie in the out of plane direction \cite{19Lohndorf} where stray field effects from domain wall formation should be observed. The measurements reported here were obtained using VSM; however SQUID measurements showed similar results. The sample dimensions were $10 \times 5$ mm$^2$.\\

\begin{figure}[h!]
\includegraphics[scale=0.36]{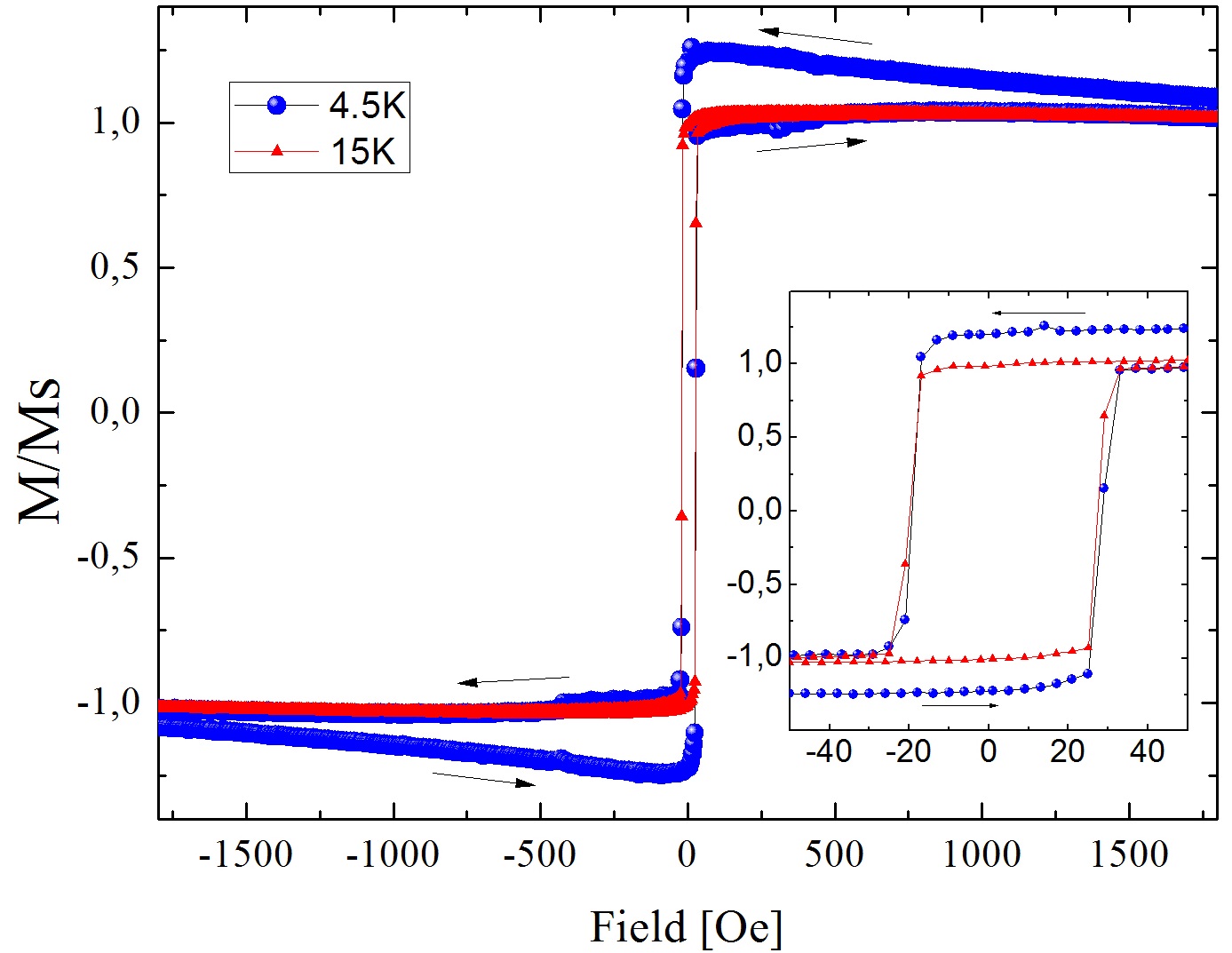}
\caption{\label{fig2}Magnetic Hysteresis loops of Nb($25$ nm)Co($45$ nm) above and below  $T_{c}$. Inset shows zoom view of hysteresis loops around coercive fields.}
\end{figure}

\begin{figure}[h!]
\includegraphics[scale=0.33]{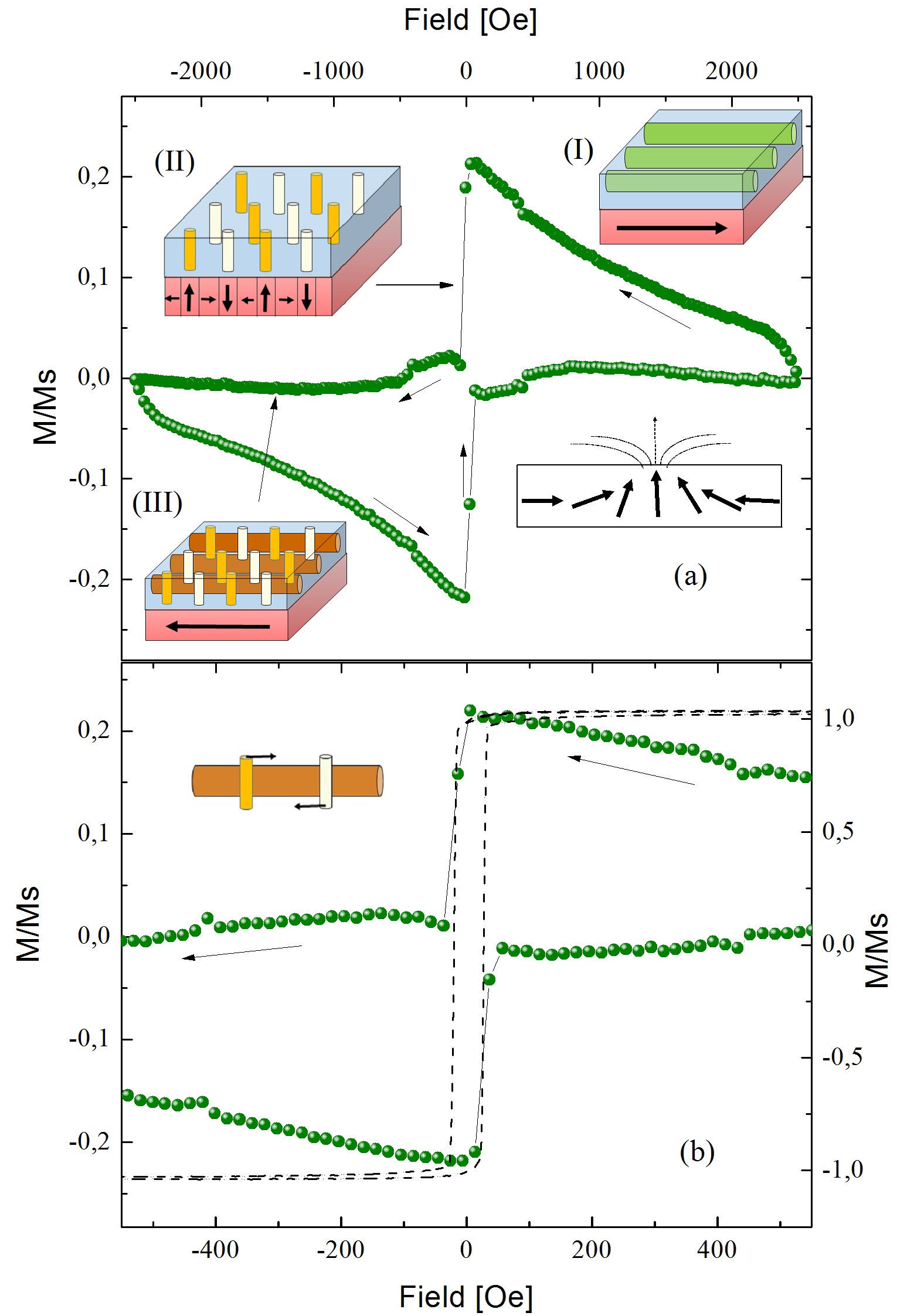}
\caption{\label{fig3}(a) Hysteresis loop corresponding to the superconducting magnetization for the Nb($25$ nm)Co($45$nm) sample along the plane. Here the magnetization at $15$ K has been subtracted from the magnetization at $4.5$ K. (b) Zoom view of the superconducting magnetization. The dotted line indicates the hysteresis loop above  $T_{c}$.}
\end{figure}

Starting at a temperature of $300$ K, the sample was cooled in zero field. In Fig. \ref{fig2} magnetization measurements of a typical Nb($25$)Co($45$) sample are shown. Here the magnetization has been normalized to its saturation value above and below $T_c \sim 6$ K.  The hysteresis loop above $T_c$ corresponds to a typical ferromagnet with in-plane magnetic anisotropy and a coercive field about $25$ Oe (see inset Fig. \ref{fig2}). Below $T_c$ the superconductor adds a magnetic contribution to the signal.

In order to clearly extract the Nb superconducting signal from the total signal, the magnetization above $T_c$ is subtracted from the magnetization below $T_c$.  The Nb signal shown in Fig. \ref{fig3}a and its zoom view in 3b is the result of subtracting the magnetization at T = $15$ K from the magnetization at T = $4.5$ K. The dotted line in Fig. \ref{fig3}b indicates the hysteresis loop above  $T_{c}$.\\

As anticipated, the presence of BDWs generates a superconductor signal which is quite different from that in a simple F/S bilayer (Fig. \ref{fig1}) without BDW. Indeed starting the analysis at high positive fields, we observe a magnetization value close to zero indicating large flux penetration. Then, decreasing the field and changing its direction at zero field, displays a magnetization increase that corresponds to vortex pinning process (I) as represented in inset of Fig. \ref{fig3}a.
This flux trapping continues up to Co coercive field, around where the signal reaches its maximal value.\\

At this field, the superconducting sample magnetization suddenly drops to nearly zero where $M$~$\sim$~$0$ and thus $B$~$\sim$~$H$ (as dictated from $B=H+4\pi M$). Given that magnetization is measured in the plane this sudden drop indicates that in-plane vortices have left the sample.\\
The data immediately after the coercive fields shows a small upturn compatible with a weak Meissner effect as result of applied field in the opposite direction. This is at first surprising given that in the absence of vortices the upturn should be enormous, similar as the one observed in Fig. \ref{fig1} after increasing the field starting from zero.  On increasing the negative field further, the magnetization remains low up to its maximum value. Finally on the return branch vortex pinning process is restored.\\

We present here a qualitative explanation for such behaviour. Around the coercive fields of the ferromagnet, the effective stray fields of Bloch domain walls are maximum, leading to an abrupt vortex penetration phase (II) in the out of plane  direction.  This effectively reduces the superconducting regions, weakening superconductivity and the pinning of the in-plane vortices. Indeed this was demonstrated in magnetoresistance measurements where a reentrant behavior to superconductivity was observed [1].  Furthermore, as follows from $M\sim 0$ at coercive fields, the weakening of superconductivity leads to a complete expulsion of in-plane vortices pinned in the sample.\\

A further increase of the external in-plane field in the negative direction will lead to elimination of BDWs disappearance as the ferromagnet saturates. Nevertheless, the superconductor's magnetic response remains low indicating BDW vortices remain trapped due to vortex pinning.\\

Eventually as the external field is increased in the negative direction, new magnetic flux should gradually be pushed into the superconductor and coexist with BDW vortices as represented in (III).  At the same time newly induced in plane vortices turn into a channel (inset Fig. \ref{fig3}b) that allow the out of plane vortices of opposite polarity to move and annihilate or leave the sample. This way the sample recovers strong superconducting properties and pinning centers become vacant to pin new in plane vortices.
Thus on the return branch the vortex pinning process is restored as shown in the experimental data.\\
The insets in Fig. \ref{fig3}a illustrate these three processes taking place; (I) vortex pinning, (II)  BDW vortex injection  and (III) vortex creation (in the plane).   \\

\begin{figure}[h!]
\includegraphics[scale=0.3]{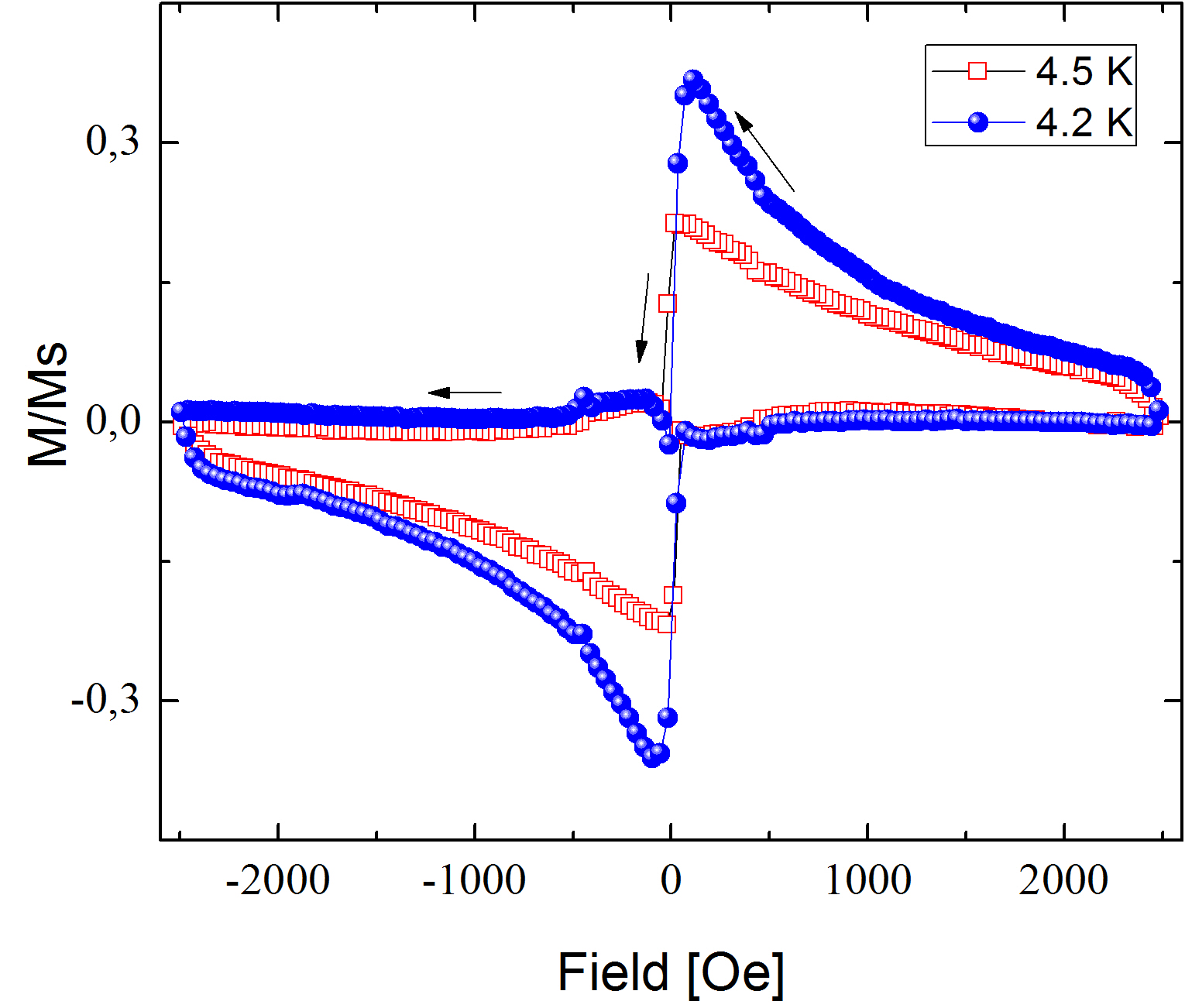}
\caption{\label{fig4}Hysteresis loop of the superconducting magnetization for Nb($25$)/Co($45$) sample (squares) and  Nb($25$)/Co($42$) (circles) taken at temperatures of $4.5$ K and $4.2$ K respectively.}
\end{figure}

Similar effects have been observed in other samples at different temperatures as shown in Fig. \ref{fig4} and summarized in Table I where magnetization drops have been found in all studied samples with Co thickness between $40$ and $55$ nm.\\

\textbf{IV. INTERPRETATION OF THE RESULTS}\\

The results presented here show sharp drops in magnetization at coercive fields. This was found in Nb/Co bilayers with Co film thicknesses range between $40$ to $55$ nm.\\

Let us compare this magnetization results with critical current measurements described in reference [1]. In this work dips in critical current measurements, that increased with ferromagnet thickness in Nb/Co bilayers were only observed beyond a thickness of the ferromagnet $d_F > 30$ nm.\\

This is of the order of thickness where domain walls of the ferromagnet, Co are thought to gradually change from being in plane (Néel domain walls) towards the out of plane direction (Bloch domain walls). For Co films, at the range of film thickness between $35$ to $100$ nm Bloch domain walls are formed with a cross-tie structure. \cite{19Lohndorf}  Here Bloch lines are located periodically with alternating polarity, pointing outwards or inwards producing high stray fields in this direction, reminiscent of a checkerboard of antiparallel stray fields.\\

In order to determine if such out of plane stray fields are strong enough to inject vortices inside the superconductor an estimation of such fields is necessary.
This can be roughly done for  Co films based on the observations reported in \cite{19Lohndorf}. Here each Bloch line has an approximate width $\delta \approx 0.15 - 0.3 \mu$m with magnetic fields $>2800$ Oe and a magnetic flux between $3-13 \ \Phi_0$, where $\Phi_0$ is the flux quantum.This is clearly much greater than $H_{c1} \sim 100$ Oe as found from our data in Fig. \ref{fig1}.
Furthermore, due to the fact that Bloch lines are antiparallel to one another a 2D vortex-antivortex array pinned by the domain walls must form as shown in the inset Fig. \ref{fig3}a. Note also that as the film gets thicker the domains reduce in size, increasing the number of domain walls and Bloch lines that can fit in the same interface area \cite{26Stankiewicz} thus enlarging the local out of plane fields on the superconductor.\\

Bloch domain wall formations in thick Co films appear to be responsible for drops in magnetization at coercive fields. Indeed for thin Co layers where Bloch domain walls are absent a full conventional type II magnetization hysteresis loop was observed (Fig. \ref{fig1}).\\

In contrast among the few reported investigations in the literature on magnetization measurements on S/F heterostructures \cite{15Patino,16Kobayashi} the measurements did not show such effect.
For the most recent study \cite{15Patino} this is easily explained as the F layers are just a few nanometers thick where no Bloch domain walls are expected. \\

On the other hand in the work by Kobayashi et al.  [16] they investigated five Nb/Co samples composed of one multilayer and four trilayers. Even though three of these had Co $\sim 50$ nm ($> 30 $nm) the authors don't report any sudden reduction in magnetization. This could be the result or significantly thicker Nb films used in this experiments with a Nb thickness of $100$ or $200$ nm. It is well known  $T_{c}$ (and thus $H_{c1}$) of the Nb films increases with thickness of the superconducting layer \cite{23Bell}.  Furthermore, given that $H_{c1}$ also reduces with temperature spontaneous vortex penetration should be more easily observable close to $T_c$ and may disappear at low temperatures.\\

\textbf{IV. SUMMARY AND CONCLUSIONS}\\

We have performed in plane magnetization measurements on Nb/Co continuous bilayer structures where the Nb $\sim 25$ nm and the Co layer has a thickness $d_{Co} \geq 40$ nm. This thick Co layers where chosen in order to investigate the influence of Bloch domain walls of the Co layer on the superconductor’s magnetization. Our results show sudden drops in the magnetization around coercive fields of the ferromagnet where $M \sim 0$ and $B\sim H$. This is the response of a normal state material with a weak or no diamagnetic response where the external field can easily penetrate. This results in a reentrant behavior to superconductivity as seen in magnetoresistance measurements in \cite{1Patino}. Similar properties of so called field-induced superconductivity (FIS) have only been realized by magnetic dots array deposited on top of a superconducting Pb film \cite{27Lange}.\\

The drops in the magnetization of the material are attributed to BDW vortex injection in the out of plane direction. The same mechanism was used to explain the drops in critical currents reported in \cite{1Patino} on very similar structures.\\
In contrast to \cite{1Patino} in this investigation there are not external currents and therefore Lorentz forces are absent. Here the magnetization drops occur at coercive fields and continues up to $H_{c2}$.\\

In summary using thin superconductor films and strong thick ferromagnets where Bloch domain walls are energetically favorable should facilitate vortex penetration in the out of plane direction.\\
This produces  two distinctive magnetic states of the superconductor i.e. with or without out of plane vortices. To inject vortices one needs a small field of about the coercive field of the ferromagnet. On the other hand to remove them a large field of the order of $H_{c2}$ is necessary.
These two well defined vortex states gives a low field switching device that could be useful for storing binary information
as superconducting memory storage devices.
\\

\textbf{ACKNOWLEDGEMENTS}\\
Edgar J. Pati\~no wishes to thank Denis Chevallier for useful comments.
\bibliographystyle{apsrev4-1}
\bibliography{Paper.bib}
\end{document}